
\def\square{\kern1pt\vbox{\hrule height 1.2pt\hbox{\vrule width 1.2pt\hskip 3pt
   \vbox{\vskip 6pt}\hskip 3pt\vrule width 0.6pt}\hrule height 0.6pt}\kern1pt}
\def\lam{{\Lambda\over 2}}
\magnification 1200

\voffset=-.1in
\vsize=7.5in
\hsize=5.6in
\tolerance 10000

\baselineskip 12pt plus 1pt minus 1pt
\pageno=0
\input font
\input rogerfonts
\centerline{\bfmone Canonical Analysis of Poincar\'e Gauge Theories}
\medskip
\centerline{\bfmone for Two Dimensional Gravity}
\smallskip
\smallskip
\vskip 24pt
\centerline{ G. Grignani}
\vskip 8pt
\centerline{\it Dipartimento di Fisica}
\centerline{\it Universit\'a degli Studi di Perugia}
\centerline{\it I-06100  Perugia -- ITALY}
\vskip 12pt
\centerline{and}
\vskip 12pt
\centerline{ G. Nardelli}
\vskip 8pt
\centerline{\it Dipartimento di Fisica}
\centerline{\it Universit\'a degli Studi di Trento}
\centerline{\it I-38050  Povo (TN) -- ITALY}
\vskip 48pt
\centerline{\bf Abstract}
\bigskip

Following the general method discussed in Refs.[1,2], Liouville gravity and
the  2 dimensional model of non-Einstenian gravity
${\cal L} \sim curv^2 + torsion^2 + cosm.\ const.$ can be formulated as
$ISO(1,1)$ gauge theories. In the first order formalism the models present,
besides the Poincar\'e gauge symmetry, additional
 local symmetries.
 We show that
in both models one can fix these additional symmetries preserving the
$ISO(1,1)$ gauge
symmetry and the diffeomorphism invariance, so that,
after a preliminary Dirac procedure, the remaining
constraints uniquely satisfy the  $ISO(1,1)$ algebra. After the
additional symmetry is fixed, the equations of motion are unaltered. One thus
remarkably simplifies the canonical structure, especially of the second model.
Moreover, one shows that the Poincar\'e group can always be used
consistently as a gauge group for gravitational theories in two dimensions.

\vfill
\vskip -12pt
\noindent  DFUPG-76-1993

\noindent UTF-292-1993
\hfill March 1993

\eject
\baselineskip 12pt plus 1pt minus 1pt

\medskip
\nobreak
As is well known, Einstein general relativity is vacuous in 2
dimensions as the Einstein tensor $G_{\mu\nu}= R_{\mu\nu} - {1\over 2} g_{\mu
\nu}R$ vanishes identically. Nevertheless, many models can be constructed to
provide,  even without matter, a substitute to general relativity [1]. In this
note  we shall deal with two of these models: Liouville gravity
[2]\footnote{$^\dagger$}{In our notation, latin indices  $a,b,c,...=0,1$
denote
internal (gauge) indices; they can be raised  and lowered with the Minkowskian
metric tensor $\eta_{ab}=\eta^{ab}={\rm  diag}(1, -1).$ The convention on the
antisymmetric symbol  is $\varepsilon^{01} =1$. Greek indices $\mu , \nu
, \rho ,... =0,1$ will denote space-time  indices.}
$$S^{(0)}_{\rm L}= \int {d^2x\over 2} \sqrt{ -g}\pi\bigl (R -\Lambda\bigr)\ \
,\eqno(1)$$ and the nameless gravity defined by the action [3]
$$S^{(0)} = \int {d^2x\over 4}\sqrt{ -g}\bigl( \gamma R^2 +
\beta { T}^a{}_{\mu\nu} { T}_a{}^{\mu\nu} +4 \Lambda\bigr)\ \ .
\eqno(2)$$
$\beta$, $\gamma$ and $\Lambda$ are constants, $\pi$ a Lagrange multiplier $R$
the scalar curvature and ${T}^a{}_{\mu\nu}$ the spacetime torsion.

According to a general procedure proposed in Ref.[4] to write any gravitational
theory with any coupling to matter  as a gauge theory of the Poincar\'e group,
 $S^{(0)}_{\rm L}$ and $S^{(0)}$ can  be replaced by equivalent
Poincar\'e gauge invariant actions. In particular, $S^{(0)}_{\rm L}$ was
studied
in Ref.[5], whereas the method outlined in Ref.[4] was applied to $S^{(0)}$
(and criticized)  in Ref.[6].

In this paper we
shall perform a canonical analysis of the Poincar\'e gauge theoretical
formulations of the two models. We shall show that the Poincar\'e gauge
symmetry is consistent with the canonical structure of both the models.
In both cases in fact one can perform a suitable $preliminary$ Dirac procedure
that allows to get rid of undesirable first class constraints (related to
$additional$ local symmetries other than the Poincar\'e one), thus obtaining a
constraint algebra that is only given by the Poincar\'e algebra.
Moreover such a Dirac procedure
preserves the diffeomorphism invariance and therefore
does not alter the equations of motion of
the original theory.
The Poincar\'e gauge theoretical formulations of
the two  models thus provides not only a unifying criterion for
treating gravitational theories, but, especially for the model (2), also a
substantial simplification of the canonical structure, contrary to
what claimed in Ref.[6].

 \medskip
The Poincar\'e
group in 2 dimensions has 3 generators, $P_a$ and $J$, associated to
translations and Lorentz transformations, respectively. They satisfy the
algebra
$$ [P_a, P_b]=0 \ , \qquad
\qquad [P_a, J]=\varepsilon_{ab} P^b \ \ .  \eqno(3)$$
 An $ISO(1,1)$ gauge
field assumes then the form
$$A_\mu \equiv A_\mu^A T_A=      e^a{}_\mu P_a + \omega_\mu J\ \
,\eqno(4)$$
where $A=(a,2)$, $T_A=(P_a,J)$ and at this stage $e^a{}_\mu=A^a{}_\mu$ and
$\omega_\mu=A^2{}_\mu$ have just the meaning of components of the gauge field
along the generators.

Under the action of the gauge group, $A_\mu$  transforms as
$$ \delta A_\mu = -\partial_\mu u - [A_\mu  , u] \equiv -
\Delta_\mu u \ \ ,  \eqno(5)$$
where $u\equiv u^A T_A=\rho^a P_a + \alpha J$ is the infinitesimal gauge
parameter. In components Eq.(5) reads
 $$\eqalignno{\delta \omega_\mu & =-\partial_\mu \alpha \ \ ,
&\hbox{(6.a)} \cr \delta e^a{}_\mu & = -\partial_\mu \rho^a
-\varepsilon^a{}_b \omega_\mu \rho^b +\alpha \varepsilon^a{}_b e^b{}_\mu
\ \ . &\hbox{(6.b)} \cr}$$

As shown in Refs.[4,5] whereas $\omega_\mu$ can be identified  with the spin
connection, $e^a{}_\mu$ cannot be viewed as the spacetime $zweibein$ if
one wants
to realize (1) and (2) as {\it Poincar\'e} gauge theories. In fact one is
forced
to introduce a set of auxiliary field $q^a(x)$ (the so called ``Poincar\'e
coordinates'') transforming as a vector under the gauge group
$$\delta q^a = \varepsilon^a{}_b\alpha q^b + \rho^a\ \ ,\eqno(7)$$
and then one has to identify the $zweibein$ $V^a{}_\mu$ with the
Poincar\'e covariant derivative of the $q^a$
$$ {\cal D}_\mu q^a (x) = \partial_\mu q^a + \omega_\mu
\varepsilon^a{}_b q^b
+e^a{}_\mu \equiv V^a{}_\mu\ \ . \eqno(8)$$
As can be easily checked from Eqs.(6) and (7),  $V^a{}_\mu$ defined as in
Eq.(8)
transforms as a Lorentz vector under Poincar\'e gauge
transformations, so that  it is quite simple to construct from Lorentz,
Poincar\'e gauge invariant actions.

Let us start our analysis from Liouville gravity, Eq.(1). For this
model,
the simplification in the canonical structure of the theory induced by our
procedure are not as drastic as for the model (2). This is related to the fact
that Liouville gravity can also be treated as a de Sitter gauge theory [7].
Nevertheless our discussion will show first, that even when other symmetries
are
present, the Poincar\'e group can be consistently used as a gauge group for
gravitational theories  and second, that there are many analogies in the
canonical analysis of  models (1) and (2).

 In the first order
(Hamiltonian) formalism, $S^{(0)}_{\rm L}$ can be replaced by the
equivalent action
$$S^{(1)}_{\rm L}=\int {d^2x\over 2} \varepsilon^{\mu\nu}
\left[\pi_a { T}^a{}_{\mu\nu} + \pi_2\left(R_{\mu\nu}+ {\Lambda\over
2}\varepsilon_{ab}e^a{}_\mu e^b{}_\nu\right)\right]\ \ ,\eqno(9)$$
where ${T}^a{}_{\mu\nu}=\partial_\mu e^a{}_\nu - \partial_\nu e^a{}_\mu +
\varepsilon^a{}_b(\omega_\mu e^b{}_\nu - \omega_\nu e^b{}_\mu)$, $R_{\mu\nu} =
\partial_\mu\omega_\nu - \partial_\nu \omega_\mu$ and $(\pi_2,\pi_a)$ are
Lagrange multipliers. Here $e^a{}_\mu$ has to be identified with the $zweibein$
and ${T}^a{}_{\mu\nu}$ with the spacetime torsion. Apart from diffeomorphisms,
$S^{(1)}_{\rm L}$ is invariant under local Lorentz transformations
$$\eqalign{\delta \pi_2&= 0\ \ ,\cr
\delta \pi_a&=\varepsilon_{ab}\alpha \pi^b\ \ ,\cr
\delta\omega_\mu&= -\partial_\mu\alpha\ \ ,\cr
\delta e^a{}_\mu &= \varepsilon^a{}_b\alpha e^b{}_\mu\ \ ,\cr}\eqno(10)$$
and under the ``enlarged translations''
$$\eqalign{\delta\pi_2 &=\varepsilon_{ab}\pi^a\kappa^b\ \ ,\cr
\delta \pi_a &= {\Lambda\over 2}\varepsilon_{ab} \kappa^b \pi_2\ \ ,\cr
\delta\omega_\mu &= {\Lambda\over 2}\varepsilon_{ab}\kappa^a e^b{}_\mu\ \ ,\cr
\delta e^a{}_\mu &= -D_\mu \kappa^a\equiv - \partial_\mu\kappa^a
-\varepsilon^a{}_b\omega_\mu\kappa^b\ \ ,\cr}\eqno(11)$$
where $\kappa^a$ is the infinitesimal local parameter of the symmetry.
The name given to the transformations (11) derives from the fact that in (11)
we have terms of pure translations [5] and terms proportional to
$\Lambda$. The transformations (10) and (11) form the de Sitter group.
The symmetry (11) appears in  $S^{(0)}_{\rm L}$ only as a dynamical symmetry.

For the specific choice $\kappa^a=e^a{}_\mu \xi^\mu$ of the
infinitesimal parameter, it can be shown that the transformations (11)
{\it on-shell} reproduce
the usual field transformations under diffeomorphisms
 $\delta x^\mu
=\xi^\mu$, up to a local Lorentz transformation with parameter  $\alpha =
\omega_\mu \xi^\mu$.
It should be remarked that to prove the equivalence between enlarged
translations and diffeomorphism transformations,
all the independent field equations have to be used in the r.h.s. of Eqs.
(11). Consequently, in this formulation of Liouville gravity, if one
breaks
the symmetry (11), necessarily  looses the general covariance of the
equations of motion.

Introducing the Poincar\'e coordinates and the definition (8) for the
$zweibein$
one can easily write from (9) the following Poincar\'e gauge invariant action
$$ S^{(2)}_{\rm L}= \int {d^2x\over 2} \, \varepsilon^{\mu \nu} \bigl[
\pi_A F^A{}_{\mu \nu}  +{\Lambda\over 2} (\pi\tilde q)
 \varepsilon_{ab} {\cal D}_\mu q^a {\cal D}_\nu q^b
\bigr] \ \ .\eqno(12)$$
Here $F_{\mu \nu}$ is the Lie algebra valued field strength
 $$\eqalign{F_{\mu \nu}&= [\Delta_\mu ,
\Delta_\nu] = P_a T^a{}_{\mu  \nu} + J R_{\mu \nu} \cr
& =P_a [\partial_\mu e^a{}_\nu - \partial_\nu e^a{}_\mu + \varepsilon^a{}_b
(\omega_\mu e^b{}_\nu - \omega_\nu e^b{}_\mu)] + J[\partial_\mu
\omega_\nu - \partial_\nu \omega_\mu]\ \ , \cr} \eqno(13)$$
transforming according to the adjoint representation of the $ISO(1,1)$ group,
whereas $\pi_A\equiv(\pi_a,\pi_2)$ is a Lagrange multiplier triplet belonging
to
the coadjoint representation
$$\eqalign{\delta\pi_2 &=\varepsilon_{ab} \pi^a\rho^b\ \ ,\cr
\delta\pi_a &=\varepsilon_{ab}\alpha\pi^b\ \ .\cr}\eqno(14)$$
 In Eq.(12) we have introduced the gauge
invariant quantity $(\pi\tilde q) = \pi_2 -\varepsilon_{ab} \pi^a q^b\equiv
\pi_A\tilde q^A,\ \tilde q^A$ being the triplet  $\tilde q^A=
(-\varepsilon^a{}_b q^b , 1)$   transforming  according to the adjoint
representation. Moreover $\sqrt{-g} = - {1\over
2}\varepsilon^{\mu\nu}\varepsilon_{ab}{\cal D}_\mu q^a {\cal D}_\nu q^b$ is
gauge invariant so that the whole action (12) is such too.
The rewriting of (9) as (12)
gives then to Liouville gravity the manifest structure of a Poincar\'e gauge
theory.

Note that in this formulation and with the identification (8) the spacetime
torsion ${\cal T}^a{}_{\mu\nu}$ is now given in terms of the field strength
components by ${\cal T}^a{}_{\mu\nu}= T^a{}_{\mu\nu} +\varepsilon^a{}_b q^b
R_{\mu\nu}$. For $q^a=0$, ${\cal T}^a{}_{\mu\nu}=T^a{}_{\mu\nu}$ and
$S^{(2)}_{\rm L}\equiv S^{(1)}_{\rm L}$.

The equations of motion deriving from (12) are
$$\eqalignno{\varepsilon^{\mu \nu} R_{\mu \nu} + {\Lambda\over 2}
\varepsilon^{\mu \nu} \varepsilon_{ab} {\cal D}_\mu q^a {\cal D}_\nu q^b
&=0 \ \ , &\hbox{(15.a)} \cr
\varepsilon^{\mu \nu} T^a{}_{\mu \nu} - {\Lambda\over 2} \varepsilon^a{}_b q^b
\varepsilon^{\mu \nu} \varepsilon_{cd} {\cal D}_\mu q^c {\cal D}_\nu q^d
&=0 \ \ , &\hbox{(15.b)} \cr
\partial_\mu \pi_2 + \varepsilon^a{}_b \pi_a e^b{}_\mu -
{\Lambda\over
2} (\pi_2 - \varepsilon^a{}_b \pi_a q^b) q_c{\cal D}_\mu q^c &=0 \ \ ,
&\hbox{(15.c)} \cr
\partial_\mu \pi_a+\varepsilon_a{}^b \omega_\mu \pi_b +
{\Lambda\over 2}\varepsilon_{ab} {\cal D}_\mu q^b (\pi_2 -\varepsilon^c{}_d
\pi_c q^d) &=0 \ \ . &\hbox{(15.d)} \cr}$$
Note  that we
have not written the equation of motion obtained variating the action with
respect to $q^a$ as redundant: it is a general feature of this formalism that
the equations for the extra degrees of freedom $q^a$ (whose introduction is
necessary in order to have a Poincar\'e gauge invariant model) always reproduce
identically satisfied conditions [4,6,8].
Eqs.(15) are easily proved to be equivalent (providing the $zweibein$
$V^a{}_\mu$ is invertible) to the equations of motion deriving from (1).

Due to the presence of the symmetry (11) in $S^{(1)}_{\rm L}$, in
$S^{(2)}_{\rm L}$ one finds, besides the Poincar\'e
and diffeomorphism invariances, an additional
non-linear local symmetry $$\eqalign{\delta_\kappa \pi_2 &=
\lam (\pi\tilde q) q_a\kappa^a\ \ ,\cr
 \delta_\kappa \pi_a &= - \lam(\pi\tilde q)\varepsilon_{ab} \kappa^b\ \ ,\cr
\delta_\kappa e^a{}_\mu &=  \lam  \varepsilon_{cd}\kappa^c {\cal
D}_\mu q^d\varepsilon^a{}_b q^b\ \ ,\cr
\delta_\kappa \omega_\mu &=  - \lam
\varepsilon_{ab} \kappa^a {\cal D}_\mu q^b\ \ ,\cr
\delta_\kappa q^a & = \kappa^a\ \ .\cr}\eqno(16)$$
 For $q^a=0$ Eqs.(16) give the part proportional to $\Lambda$ in (11) up to an
overall sign.

{}From (16) one sees that introducing the Poincar\'e coordinates, so as to
realize the model as a Poincar\'e gauge theory, one has decoupled the
translations and the transformations proportional to $\Lambda$ in (11), so
that (12) is independently invariant under these transformations.
Similarly to what  happened in the previous formulation of Liouville gravity,
by choosing $\kappa^a = {\cal D}_\mu q^a \xi^\mu$  Eqs. (16) are easily
seen to reproduce
{\it on-shell} diffeomorphism transformations $\delta_\xi$,
 up to local Poincar\'e  gauge transformations
$\delta_P$ with
parameters $\rho^a =e^a{}_\mu \xi^\mu$ and $\alpha =\omega_\mu \xi^\mu$.
Consequently, the symmery (16) can be rewritten as
$\delta_\kappa = \delta_\xi + \delta_P + \delta_{OS}$, where
$\delta_{OS}$ is a symmetry of the action (12) that vanishes {\it
on-shell}, and therefore proportional to the equations of motion.

We are now ready to perform the canonical analysis. Having 5 local
symmetries in the action $S^{(2)}_{\rm L}$ we expect to find 5 first class
constraints.

To make manifest the canonical structure of the theory and to simplify the
generator algebra it is convenient to consider instead of (12) the
equivalent action
$$S=\int d^2x \left(p_a \dot q^a + \pi_2 \dot \omega_1 +\pi_a \dot e^a{}_1
+\lambda^a J_a + \omega_0 G_2 + e^a{}_0 G_a\right)\ \ .\eqno(17)$$
Here $\lambda^a$ is a Lagrange multiplier transforming as a Lorentz vector
under $ISO(1,1)$ gauge transformations and
$$\eqalignno{
G_2&= \partial_1
\pi_2 + \varepsilon_{ab} \pi^a e^b{}_1 +\varepsilon_{ab}p^aq^b\ \ ,
&\hbox{(18.a)} \cr
G_a&= \partial_1 \pi_a +\varepsilon_{ab} \pi^b \omega_1 + p_a\ \ ,
&\hbox{(18.b)}\cr J_a&= p_a - \lam (\pi\tilde q) \varepsilon_{ab}
{\cal D}_1 q^b\
\ , &\hbox{(18.c)}  \cr}$$
are the generators of the 5 local symmetries. The equivalence of (17-18) and
(12) is obvious since eliminating $\lambda^a$ and $p_a$ in (17) one obtains
(12).

As $G_2$ and $G_a$ are multiplied in (17) by the zero components of the
gauge potential $A_0=(e^a{}_0,\omega_0)$, they are the generators of the
Poincar\'e gauge symmetry and, as we shall see, the $J_a$ are the generators of
the additional symmetry (16).

{}From (17) one sees that $\pi_A =(\pi_a, \pi_2)$ are the momenta
canonically conjugate to $A_1^A = (e^a{}_1 ,  \omega_1)$.
Since  the time derivatives of  $A_0^A = (e^a{}_0 ,  \omega_0)$
and $\lambda_a$
do not appear in (17), the definition of their conjugate momenta
leads to 5 primary constraints
$$\eqalignno{\pi^{(0)}_2 & = {\partial L \over \partial
\dot\omega_0}  \simeq 0\ \ , &\hbox{(19.a)}\cr
\pi^{(0)}_a & = {\partial L \over \partial \dot e^a{}_0}
\simeq 0\ \ , &\hbox{(19.b)}\cr
\pi^{(\lambda)}_a & = {\partial L \over \partial \dot \lambda^a}
\simeq 0 \ \ .  &\hbox{(19.c)}\cr}$$
The symbol ``$\simeq$''  denotes ``weakly equal'' in
the Dirac terminology, namely the constraints (19.a -- 19.c)
 have non-vanishing canonical
brackets with some canonical variables and are then incompatible with the
canonical Poisson brackets [9]. The non-vanishing fundamental Poisson brackets
are
 $$\eqalign{\{e^a{}_0(x), \pi^{(0)}_b(y)\} &=\delta^a{}_b \delta (x-y)\ \ ,\cr
\{e^a{}_1(x), \pi_b(y)\} &=\delta^a{}_b \delta (x-y)\ \ ,\cr
\{\omega_0(x), \pi^{(0)}_2(y)\} &= \delta (x-y)\ \ ,\cr
\{\omega_1(x), \pi_2(y)\} &= \delta (x-y)\ \ ,\cr
\{q^a(x), p_b(y)\} &=\delta^a{}_b \delta (x-y)\ \ ,\cr
\{\lambda^a (x) ,\pi^{(\lambda)}_b (y)\}
&=\delta^a{}_b \delta (x-y)\ \ . \cr}\eqno(20)$$
The canonical Hamiltonian ${H}$ one gets from Eqs. (17 -- 19) is
 $$ {H}=\int dx {\cal H} =\int dx \left(\pi^{(0)}{}_a \dot
e^a{}_0 + \pi^{(0)}{}_2 \dot \omega_0
+\pi^{(\lambda)}{}_a \dot \lambda^a
 -\lambda^a J_a
 - \omega_0 G_2 - e^a{}_0 G_a\right)
\ \ ,\eqno(21)$$
and $H$ is completely expressed in terms of constraints as one expects in a
theory of gravity.

 Using (20) and (21) one can impose the temporal consistency of
the primary constraints (19). This gives rise to the secondary constraints
$G_a\simeq 0$, $G_2\simeq 0$ and $J_a\simeq 0$. The ``Gauss's laws'' related to
the gauge symmetry are, as expected the $G_A\simeq 0$.
The algebra satisfied by the generators can be obtained using (20) and for the
$G_A$ is given by the Poincar\'e algebra (3)
$$ \{ G_a (x), G_b (y)\} =0 \quad, \qquad
\{ G_a (x), G_2 (y)\} =\varepsilon_{ab} G^b  \delta (x-y)
\ \ .\eqno(22)$$
The Poisson brackets involving the
 constraints $J_a$ are
$$\eqalignno{ \{G_a (x), J_b(y)\} &=0\ \ , &\hbox{(23.a)}\cr
\{J_a (x), G_2(y)\} &=\varepsilon_{ab} J^b \delta(x-y)\ \ , &\hbox{(23.b)}\cr
\{J_a (x), J_b(y)\} &=
\varepsilon_{ab}\lam (G\tilde q) \delta(x-y)\ \ , &\hbox{(23.c)}\cr}$$
where $(G\tilde q)= G_A\tilde
q^A=G_2-\varepsilon_{ab}G^aq^b$. Thus all the constraints are
first class having Poisson brackets among themselves that weakly vanish. In the
$\{J_a, J_b\}$ bracket one finds ``structure functions'' (depending
on the $q^a$) rather than structure constants.

The existence of first class
constraints is always related to the presence of a local symmetry.
The functional  ${\cal G} =\int dx (\rho^a G_a +\alpha G_2)$
obviously generate
the Poincar\'e transformations, whereas ${\cal J}
=\int dx \kappa^a J_a$ provides the following transformations of the canonical
variables
$$\eqalign{\delta_\kappa \pi_2 &\equiv\{\pi_2,{\cal J}\}=  \lam
(\pi\tilde q) q_a\kappa^a\ \ ,\cr
 \delta_\kappa \pi_a & \equiv\{\pi_a,{\cal J}\}= - \lam(\pi\tilde q)
\varepsilon_{ab}
\kappa^b\ \ ,\cr
\delta_\kappa e^a{}_1 &\equiv\{e^a{}_1,{\cal J}\}=\lam
\varepsilon_{cd}\kappa^c
{\cal D}_1 q^d\varepsilon^a{}_b q^b\ \ ,\cr
\delta_\kappa \omega_1 &\equiv\{\omega_1,{\cal J}\}= - \lam
\varepsilon_{ab} \kappa^a {\cal D}_1 q^b\ \ ,\cr
\delta_\kappa q^a &\equiv\{q^a,{\cal J}\} = \kappa^a\ \ ,\cr
\delta_\kappa p_a &\equiv\{p_a,{\cal J}\} = \varepsilon_{ab}\left[
\pi^b\varepsilon_{cd}\kappa^c{\cal D}_1 q^d + D_1\left((\pi\tilde
q)\kappa^b\right)\right]\ \ .\cr}\eqno(24)$$
A comparison with Eqs. (16) shows that the $J_a$ are indeed the
canonical generators of the non-linear symmetry (16).
As usual in a Hamiltonian
formulation, the transformations of the Lagrange multipliers are instead
obtained
by consistency of the transformation laws of the canonical variables with the
equations of motion. With this procedure we get
$$\eqalign{\delta_\kappa\lambda^a &= - D_0\kappa^a\ \ ,\cr \delta_\kappa
e^a{}_0
&= -\lam
  \varepsilon_{cd}\kappa^c\lambda^d\varepsilon^a{}_b q^b\ \ ,\cr
\delta_\kappa\omega_0 &=  \lam\varepsilon_{ab}\kappa^a\lambda^b\ \ ,\cr}
\eqno(25)$$
and Eqs.(24-25) provide a local invariance of the action (17-18).

The introduction of the Poincar\'e coordinates and the consequent decoupling of
the translations and of the $J$-symmetry allows a reduction
 of the constraint algebra, Eqs.(22-23), to the sole Poincar\'e algebra.
In fact one can now fix the $J$-symmetry, so as to eliminate the $J_a$
constraints, preserving  the Poincar\'e gauge invariance
and  the general covariance of the
 equations of motion.
This is an important difference with the previous formulation of
Liouville gravity, Eq. (9), where a breaking of the symmetry (11) was
associated  to a loss of general covariance in the field equations.
One can do
so by imposing the gauge $covariant$ auxiliary condition
$$\sigma_a=c\lambda_a - \pi_a\ \ ,\eqno(26)$$
where $c$ is an arbitrary constant
with dimensions $length^{-1}$ [$c$ can be chosen for example as
$c=\sqrt{|\Lambda|}$, however, since the specific value of $c$ is completely
irrelevant, we shall leave it arbitrary].
The choice (26) is always possible being now $\lambda_a$ a dynamical variable
(see (21)). Note that the possibility of choosing the constraint (26) arise
only in the Poincar\'e gauge theoretical approach; as a matter of fact  the
introduction of $\lambda_a$ is forced in the Dirac-Hamiltonian analysis of the
Poincar\'e gauge theory for Liouville gravity (12), because $\lambda_a$ is the
Lagrange multiplier of the primary constraint deriving from the
definition of the
momentum conjugate to $q^a$. Thus  the introduction of the Poincar\'e
coordinates $q^a$ unavoidably leads to the introduction of the $\lambda_a$.

The constraints (26) allow to get rid of the $J_a=0$ constraints making them
second class. In fact the matrix of the Poisson brackets $\{\phi_\alpha,
\phi_\beta\}$, where $\phi_\alpha\equiv(\sigma_a,\ J_a)$, is non-singular.
Consequently, one can perform a preliminary Dirac procedure with these
constraints to construct canonical brackets that are consistent with the
setting of the constraints $J_a$ and $\sigma_a$ strongly to zero.

The matrix of the
Poisson brackets $\{\phi_\alpha , \phi_\beta\}$ is
 given by
$$C_{\alpha \beta} (x,y)= \{\phi_\alpha (x) , \phi_\beta (y)\}=
\lam (\pi\tilde q)
\left(\matrix{0&\varepsilon_{ab}\cr\varepsilon_{ab}& 0\cr}\right)
\delta(x-y)\ \ ,\eqno(27)$$
so that its inverse is
$$[C^{-1} (x,y)]^{\alpha\beta} =\left[\lam(\pi\tilde q)\right]^{-1}
\left(\matrix{0&\varepsilon^{ab}\cr\varepsilon^{ab}& 0\cr}\right)
\delta(x-y)\ \ .\eqno(28)$$
The Dirac brackets for any pair  ${\cal A}(x), {\cal B}(y)$
of functionals  of canonical variables read
$$\eqalign{ \{ {\cal A}(x), {\cal B}(y)\}_{\cal D} & =
\{ {\cal A}(x), {\cal B}(y)\} - \int du \,
\{ {\cal A}(x), \sigma_a(u)\}  {2\varepsilon^{ab}\over\Lambda(\pi\tilde q)(u)}
 \{ J_b(u), {\cal B}(y)\}\cr & - \int du \,
\{ {\cal A}(x), J_a(u)\}  {2\varepsilon^{ab}\over \Lambda(\pi\tilde q)(u)}
 \{ \sigma_b(u), {\cal B}(y)\}\ \ .}
\eqno(29)$$

 The constraint algebra in terms of Dirac
brackets then becomes
$$ \{ G_a (x), G_b (y)\}_{\cal D} =0 \quad, \qquad
\{ G_a (x), G_2 (y)\}_{\cal D} =\varepsilon_{ab} G^b \delta (x-y)
\ \ ,\eqno(30)$$
all  the brackets involving the $J_a$ and the $\sigma_a$ being strongly zero.

Thus one is left with the sole Poincar\'e algebra.
This is a consequence of the fact that the auxiliary
conditions $\sigma_a$ do not entail any
violation of the gauge symmetry.

The equations of motion
generated by the brackets (29) with the Hamiltonian $H$ are
$$\eqalignno{ \dot q^a &=\{ q^a , H\}_{\cal D} =
-\varepsilon^a{}_b q^b \omega_0 -e^a{}_0 - {2\over \Lambda ( \pi\tilde
q)}\varepsilon^a{}_b D_0 \pi^b\ \ , &\hbox{(31.a)}\cr
\dot e^a{}_1 &=\{ e^a{}_1 , H\}_{\cal D} =
\partial_1 e^a{}_0 -\varepsilon^a{}_b (\omega_0 e^b{}_1 - \omega_1
e^b{}_0) + \varepsilon_{cd}{\cal D}_0 q^c {\cal D}_1 q^d
\lam(\pi\tilde q)\varepsilon^a{}_b q^b\ , &\hbox{(31.b)}\cr
\dot \omega_1 &=\{ \omega_1 , H\}_{\cal D} =
\partial_1 \omega_0 - \lam (\pi \tilde q)
\varepsilon_{cd}{\cal D}_0 q^c {\cal D}_1 q^d\ \ , &\hbox{(31.c)}\cr
\dot \pi_2 &=\{ \pi_2 , H\}_{\cal D} =
\varepsilon_{ab}  e^a{}_0 \pi^b+ \lam (\pi\tilde q) q_a {\cal D}_0q^a\ \ .
&\hbox{(31.d)} \cr}$$
Eqs. (31) are absolutely {\it identical} to  Eqs.(15), and therefore
are general
covariant.
 The mechanism that permits to break the
$J$-symmetry preserving the Poincar\'e gauge symmetry and maintaining the
diffeomorphism invariance
is  related to the fact that, within our formalism, a variation
of the action (12) with respect to the Poincar\'e coordinates $q^a$ always
gives  ``redundant'' field equations.
Recalling  the decomposition
$\delta_\kappa = \delta_\xi + \delta_P + \delta_{OS} $
of the symmetry (16),
the presence of extra-field equations (which are automatically
satisfied)  allows
to break only the $\delta_{OS}$ part of the $\delta_\kappa$ symmetry,
without spoiling neither $\delta_\xi$ nor $\delta_P$.
In fact, from Eq.  (31.a) one sees that the equation of motion for
$q^a$, that before was identically satisfied, has now become one of the
independent field equations of the theory. Namely the additional
constraints $\sigma^a$ have
projected the set of field equations into the set of
{\it independent} field equations.
The same arguments do not apply to the formulation (9) of
Liouville gravity, as all the field equations are already independent.

By choosing, instead of $\sigma_a =0$, a gauge condition that breaks the
 Poincar\'e symmetry one alters the equations of motion.
Nevertheless, if one breaks only the translational part of the gauge
symmetry (so that one is left only with the local Lorentz invariance) the
equations of motion, even if different from Eqs.(15), must still be
equivalent to those obtained variating $S^{(0)}_{\rm L}$. For
example, with the physical gauge $q^a=0$  one fixes
the translations and therefore one makes second
class the constraints $G_a$ (or the $J_a$). The remaining generators $J_a$ (or
the $G_a$) and $G_2$ and the algebra become those of the de Sitter gauge
theoretical formulation of Liouville gravity [7]. The equations of motion are
Eqs.(15) with $q^a=0$.

 Even if our treatment has not produced a simplification in the
canonical structure of Liouville gravity (one can argue whether the Poincar\'e
algebra is simpler than the de Sitter one) it has shown that the Poincar\'e
group can be used consistently as a gauge group for this theory.

Let us now turn to the model (2). The canonical analysis can be
performed exactly in the same way. The first order formalism for (2) is
provided
by the action $$\eqalign{S^{(1)}&=\int d^2x {\cal L}^{(1)}=\int {d^2x\over 2}
\varepsilon^{\mu\nu}  \left(\pi_a { T}^a{}_{\mu\nu} + \pi_2 R_{\mu\nu}+ E
\varepsilon_{ab}e^a{}_\mu
e^b{}_\nu\right)\ \ ,\cr
E &= {(\pi_2)^2\over 4\gamma} -{\pi_a\pi^a\over 2\beta} -\Lambda\ \
.\cr}\eqno(32)$$
 $S^{(1)}$ becomes $S^{(0)}$ when the equations
of motion for $\pi_a$ and $\pi_2$ are substituted back in (32). The action
$S^{(1)}$ has, besides the local Lorentz and diffeomorphisms
invariances, the additional non linear local symmetry
$$\eqalign{\delta_\kappa \pi_2 &= \varepsilon_{ab} \pi^a \kappa^b\ \ ,\cr
\delta_\kappa \pi_a &= E\varepsilon_{ab} \kappa^b\ \ ,\cr
\delta_\kappa e^a{}_\mu &= - D_\mu \kappa^a -
{1\over \beta} \pi^a \varepsilon_{bc} \kappa^b
e^c{}_\mu\ \ ,\cr
\delta_\kappa \omega_\mu &=
{\pi_2\over 2\gamma} \varepsilon_{ab} \kappa^a e^b{}_\mu\ \ ,\cr}\eqno(33)$$
Here again we have translations plus transformations (proportional to $E$,
$1/\beta$ and $1/\gamma$) that are non linear or depend on $\Lambda$.
The invariance of $S^{(1)}$ under (33) is due to the fact that
$\delta_\kappa{\cal L}^{(1)}=
\varepsilon^{\mu\nu}\varepsilon_{ab}\partial_\mu[(E+ 2\Lambda)\kappa^a
e^b{}_\nu]$.
Such a symmetry appears in the model described by (2) only as a dynamical
symmetry (at the Hamiltonian level) and in this framework it has been
extensively discussed in Ref.[10].
As it happened in the formulation (9) of Liouville gravity,
 by choosing $\kappa^a=e^a{}_\mu\xi^\mu$,
the transformations (33) {\it on-shell} reproduce diffeomorphism
transformations $\delta_\xi$, up to local Lorentz transformations
$\delta_L$ [11].

Following the recipe,
one can easily write an $ISO(1,1)$ gauge invariant action
${S}^{(2)} $ that is classically equivalent to $S^{(1)}$ and is
given by
$$ {S}^{(2)} =\int d^2x {\cal L}^{(2)}= \int
{d^2x\over 2} \varepsilon^{\mu\nu}\left(\pi_A F^A{}_{\mu\nu} + \varepsilon_{ab}
{\cal D}_\mu q^a  {\cal D}_\nu q^b \tilde E\right)\ \ , \eqno(34.a)$$
where
$$\tilde E  =  {1\over 4\gamma} (\pi \tilde q)^2
 -{1\over 2\beta} \pi_a \pi^a -\Lambda
\ \ , \eqno(34.b)$$
is a Poincar\'e gauge invariant quantity.
 ${ S}^{(2)}$ is also invariant under diffeomorphisms and under the
following local transformations (the equivalent of (16) for Liouville gravity)
$$\eqalign{\delta_\kappa \pi_2 &=
\tilde E q_a\kappa^a\ \ ,\cr
 \delta_\kappa \pi_a &= -\tilde E\varepsilon_{ab} \kappa^b\ \ ,\cr
\delta_\kappa e^a{}_\mu &= \varepsilon_{cd}\kappa^c {\cal D}_\mu
q^d\left({(\pi\tilde q)\over 2\gamma}\varepsilon^a{}_b q^b +{1\over
\beta}\pi^a\right)\ \ ,\cr
\delta_\kappa \omega_\mu &=  -{(\pi\tilde q)\over 2\gamma}
\varepsilon_{ab} \kappa^a {\cal D}_\mu q^b\ \ ,\cr
\delta_\kappa q^a & = \kappa^a\ \ .
\cr}\eqno(35)$$
In fact
$\delta_\kappa{\cal L}^{(2)}= -
\varepsilon^{\mu\nu}\varepsilon_{ab}\partial_\mu[(\tilde E+2\Lambda)\kappa^a
{\cal D}_\nu q^b]$. Note that for $q^a=0$, as expected, Eqs.(35) provide
the non linear and the $\Lambda$ parts of Eqs.(33), up to a global sign.
Again, by choosing $\kappa^a = {\cal D}_\mu q^a \xi^\mu$,
Eqs.  (35) {\it on-shell} can be decomposed as
the sum of diffeomorphisms
 plus Poincar\'e gauge transformations
$\rho^a = e^a{}_\mu \xi^\mu$ and $\alpha = \omega_\mu \xi^\mu$.

The  independent equations of motion are those obtained variating the action
$S^{(2)}$ with respect to
$\pi_2 , \ \pi_a , \ \omega_\mu$ and  $e^a{}_\mu$, and
are given by, respectively,
$$\eqalignno{R_{01} + {(\pi \tilde q)\over 2\gamma} \varepsilon_{ab}
{\cal D}_0 q^a
{\cal D}_1 q^b &=0\ \ , &\hbox{(36.a)}\cr
T^a{}_{01} -\left[ {1\over 2\gamma}(\pi \tilde q) \varepsilon^a{}_bq^b
+{1\over \beta}\pi^a\right]
\varepsilon_{cd}{\cal D}_0 q^c
{\cal D}_1 q^d &=0\ \ , &\hbox{(36.b)}\cr
\partial_\mu\pi_2 +\varepsilon_{ab} \pi^a e^b{}_\mu - q_a{\cal D}_\mu
q^a \tilde E &=0\ \ , &\hbox{(36.c)}\cr
\partial_\mu\pi_a +\varepsilon_{ab} \pi^b \omega_\mu
+\varepsilon_{ab}{\cal D}_\mu
q^b \tilde E &=0\ \ . &\hbox{(36.d)}\cr}
$$

As before it is
convenient to consider instead of (34) the equivalent action (17) where the
$G_a$ and $G_2$ are still given by Eqs.(18.a,b) whereas the $J_a$ become
$$J_a= p_a - \tilde E \varepsilon_{ab} {\cal D}_1 q^b\ \ .
\eqno(37)$$

The primary constraints, the non-vanishing fundamental Poisson brackets and the
canonical Hamiltonian, are provided by  (19) (20) and (21)
respectively. The Gauss's laws are $G_A\simeq 0$, so
that the only difference with Liouville gravity  is in the secondary
constraints
$J_a\simeq 0$, because the $J_a$ are now given by Eq.(37).

The $G_A$ obviously  satisfy the Poincar\'e algebra and the only bracket
that changes in Eqs.(23) is (23c) that now reads
$$\{J_a (x), J_b(y)\} =
\varepsilon_{ab}\left[{1\over 2\gamma}(\pi\tilde q) (G\tilde q) -{1\over
\beta} \pi^c (G_c -J_c)\right]\delta(x-y)\ \ , \eqno(38)$$
Contrary to what happens in Liouville gravity, even when
$q_a=0$
 the algebra of the $J_a$ generators
is non linear and contains structure functions instead of constants.
 This is due to the fact that the secondary constraints arising from the
action (32) ($i.e.$ Eq.(34) with $q^a=0$) are
$$K_a = G_a - J_a\bigl|_{q^a=0}\simeq0\ ,\quad\quad G_2\bigl|_{q^a=0}\simeq
0\ \ ,$$  and their Poisson algebra is  still quite
nasty. The functional ${\cal K}=\int dx \kappa^a K_a$ generates in fact the
non-linear symmetry (33) and not a honest gauge symmetry as it happens in
Liouville gravity (see Eq.(11)). Consequently, the elimination of the
constraints $J_a$ (or the $K_a$) while preserving invariance under general
coordinate transformations
(which is impossible in the formulation (32)),  now becomes  much
more important in order to simplify the Hamiltonian structure of the model.

The functional ${\cal J}=\int dx J_a\kappa^a$
now generates the following transformations on the canonical variables
$$\eqalign{\delta_\kappa \pi_2 &\equiv\{\pi_2,{\cal J}\}=
\tilde E q_a\kappa^a\ \ ,\cr
 \delta_\kappa \pi_a & \equiv\{\pi_a,{\cal J}\}=-\tilde E\varepsilon_{ab}
\kappa^b\ \ ,\cr
\delta_\kappa e^a{}_1 &\equiv\{e^a{}_1,{\cal J}\}=  \varepsilon_{cd}\kappa^c
{\cal D}_1 q^d\left[{(\pi\tilde q)\over 2\gamma}\varepsilon^a{}_b q^b
+{\pi^a\over \beta}\right]\ \ ,\cr
\delta_\kappa \omega_1
&\equiv\{\omega_1,{\cal J}\}= - {(\pi\tilde q)\over 2\gamma}
\varepsilon_{ab} \kappa^a {\cal D}_1 q^b\ \,\cr
\delta_\kappa q^a &\equiv\{q^a,{\cal J}\} = \kappa^a\ \ ,\cr
\delta_\kappa p_a &\equiv\{p_a,{\cal J}\}= \varepsilon_{ab}\left[{(\pi\tilde
q)\over 2\gamma}\pi^b \varepsilon_{cd}\kappa^c{\cal D}_1 q^d + D_1\left(\tilde
E\kappa^b\right)\right]\ \ ,\cr}\eqno(39)$$
namely the $J_a$ constraints are the canonical generators associated to
the symmetry (35).
For consistency of (39) with the
equations of motion one then finds the transformation laws of the Lagrange
multipliers
$$\eqalign{\delta_\kappa\lambda^a &=- D_0\kappa^a+
\varepsilon_{bc}\kappa^b\lambda^c {\pi^a\over\beta} \ \ ,\cr
\delta_\kappa e^a{}_0
&= - \varepsilon_{cd}\kappa^c
\lambda^d \left[{(\pi\tilde q)\over 2\gamma}\varepsilon^a{}_b q^b
+{\pi^a\over \beta}\right]\ \ ,\cr
\delta_\kappa\omega_0 &={(\pi\tilde q)\over 2\gamma}
\varepsilon_{ab} \kappa^a \lambda^b\ \ ,\cr}
\eqno(40)$$
so that Eqs.(39-40) provide the local invariance of the action (17) with
$G_a$ and $J_a$ given by (18.a,b) and (37).

Even in this case, with the same procedure, one can fix the $J$-symmetry
without breaking neither the  Poincar\'e gauge symmetry
nor the diffeomorphism invariance by introducing the
additional constraints $\sigma_a\simeq0$, Eq.(26).

The matrix of the
Poisson brackets $\{\phi_\alpha , \phi_\beta\}$ is non-singular and
is
 given by
$$C_{\alpha \beta} (x,y)= \{\phi_\alpha (x) , \phi_\beta (y)\}=\tilde E
\left(\matrix{0&\varepsilon_{ab}\cr\varepsilon_{ab}& 0\cr}\right)
\delta(x-y)\ \ ,\eqno(41)$$
the inverse being
$$[C^{-1} (x,y)]^{\alpha\beta} =\tilde E^{-1}
\left(\matrix{0&\varepsilon^{ab}\cr\varepsilon^{ab}& 0\cr}\right)
\delta(x-y)\ \ ,\eqno(42)$$
so that Dirac brackets compatible with the constraints $\phi_\alpha $
strongly set to zero,  can be
consistently defined and read
$$\eqalign{ \{ {\cal A}(x), {\cal B}(y)\}_{\cal D} & =
\{ {\cal A}(x), {\cal B}(y)\} - \int du \,
\{ {\cal A}(x), \sigma_a(u)\}  {\varepsilon^{ab}\over \tilde E(u)}
 \{ J_b(u), {\cal B}(y)\}\cr & - \int du \,
\{ {\cal A}(x), J_a(u)\}  {\varepsilon^{ab}\over \tilde E(u)}
 \{ \sigma_b(u), {\cal B}(y)\}\ \ .}
\eqno(43)$$

 The constraint algebra in terms of Dirac
brackets is solely given by the Poincar\'e algebra (30)
and the equations of motion
$$\eqalignno{ \dot q^a &=\{ q^a , \tilde H\}_{\cal D} =
-\varepsilon^a{}_b q^b \omega_0 -e^a{}_0 - {1\over \tilde
E}\varepsilon^a{}_b D_0 \pi^b &\hbox{(44.a)}\cr
\dot e^a{}_1 &=\{ e^a{}_1 , \tilde H\}_{\cal D} =
\partial_1 e^a{}_0 -\varepsilon^a{}_b (\omega_0 e^b{}_1 - \omega_1
e^b{}_0) + \varepsilon_{cd}{\cal D}_0 q^c {\cal D}_1 q^d
\left[{1\over \beta} \pi^a + {1\over 2\gamma}(\pi\tilde
q)\varepsilon^a{}_b q^b \right] &\hbox{(44.b)}\cr
\dot \omega_1 &=\{ \omega_1 , \tilde H\}_{\cal D} =
\partial_1 \omega_0 - {1\over 2\gamma} (\pi \tilde q)
\varepsilon_{cd}{\cal D}_0 q^c {\cal D}_1 q^d &\hbox{(44.c)}\cr
\dot \pi_2 &=\{ \pi_2 , \tilde H\}_{\cal D} =
\varepsilon_a{}^b \pi_b e^a{}_0 + \tilde E q_a {\cal D}_0q^a
&\hbox{(44.d)} \cr}$$
are again  {\it identical} to the equations of motion
obtained from the action (35), Eqs. (36).
On the contrary,
 if  we tried to eliminate the non-linear symmetry
without introducing the Poincar\'e coordinates,
   we would have
unavoidably spoiled
the general covariance of the equations of motion.

Therefore, thanks to
our method for writing the model as a gauge theory, even in this case one can
eliminate the  $J_a=0$ constraints preserving the invariance under general
coordinate transformations
and ending up with  a constraint algebra
that is a representation of the Lie algebra of the gauge group.

Since in this model
the algebra of the $J_a$ generators contains structure functions
instead of constants even when $q^a=0$, the
simplification induced by the Poincar\'e gauge invariant formalism is then
 quite remarkable   in this case.
This can be easily realized  by choosing the 	``physical gauge''
$q^a=0$
instead of the condition (26).
With this choice one fixes the translational part of
the Poincar\'e symmetry, and therefore makes second class the
constraints $G_a$ (or the $J_a$). The remaining generators $J_a$ (or
$G_a$) and $G_2$ reproduce the generators of the non-linear part of the
symmetry (33) plus Lorentz ($G_2$) transformations, and one returns to
the original model, Eq. (32). In doing this, however, the gauge symmetry
is lost and one is left with an undesirable constraint algebra ({\it
i.e.} Eq.
(38) with $q^a=0$).

In conclusion, with our procedure, whatever is the ``natural'' symmetry of the
model under consideration, without altering the equations of motion, one can
always end up with the sole constraints generating the gauge symmetry and thus
satisfying an algebra that is a representation of the Lie algebra
of the gauge group .
Consequently, the constraint algebra does not contain structure functions
(one of
the characteristic difficulties of gravitational theories), but plain structure
constants and one can then start to face quantization.

 \vfill
\eject

\bigskip
\bigskip

\noindent{\bf REFERENCES}
\medskip
\nobreak
\bigskip
\item{[1]} R. Mann, in {\it Proceedings of the Fourth
Canadian Conference on General Relativity and Relativistic
Astrophysics}, to be published. See also A. H. Chamseddine, {\it Phys.
Lett.} {\bf B256}, (1991) 379; I. M. Lichtzier and S. D. Odintsov, {\it
Mod. Phys. Lett.} {\bf A6}, (1991) 1953.
\medskip
\item{[2]}C. Teitelboim, {\it Phys. Lett. } {\bf B126},  (1983) 41, and
in {\it Quantum Theory of Gravity}, S Christensen ed. (Adam Higler,
Bristol, 1984); R. Jackiw in
{\it Quantum Theory of Gravity}, S Christensen ed. (Adam Higler,
Bristol, 1984) and {\it Nucl. Phys. } {\bf B252}, (1985) 343.
\medskip
\item{[3]} M. O. Katanaev and I. V. Volovich, {\it Ann. Phys.} (N.Y.)
 {\bf 197}, (1986) 1.
\medskip
\item{[4]} G. Grignani and G. Nardelli, {\it Phys. Rev. } {\bf D45},
(1992) 2719
 \medskip
\item{[5]}  G. Grignani and G. Nardelli, ``Poincar\'e Gauge Theories for
Lineal Gravities'',  Preprint
DFUPG-57-1992/UTF-266-1992.
\medskip
\item{[6]} T. Strobl, ``Comment on Gravity and the Poincar\'e Group'',
Preprint TUW-01-1993.
\medskip
\item{[7]} T. Fukuyama and K. Kamimura, {\it Phys. Lett.} {\bf 160B},
 (1985) 259; K. Isler and C. Trugenberger, {\it Phys. Rev. Lett.} {\bf
63}, (1989) 834; A. Chamseddine and D. Wyler, {\it Phys. Lett.} {\bf
B228},  (1989) 75.
\medskip
\item{[8]}D. Cangemi and R. Jackiw, {\it Ann. Phys.} (N.Y.) (in press),
MIT-CTP preprint \# 2165.
\medskip
\item{[9]} See for example A. Hanson, T. Regge and C. Teitelboim,
``Constrained Hamiltonian Systems'', Accademia Nazionale dei Lincei,
Roma 1979
\medskip
\item{[10]}H. Grosse, W. Kummer, P. Presnaider and D. J. Schwarz,
{\it J. Math. Phys.} {\bf 33}, (1992) 3892.
\medskip
\item{[11]}T. Strobl, {\it Int. J. Mod. Phys.} {\bf 8},  (1993) 1883.

\vfill
\eject

\end